\documentclass[prl,aps,a4paper,showpacs,twocolumn,floatfix]{revtex4}

\usepackage{graphicx}
\usepackage[ansinew]{inputenc}
\usepackage{array}
\usepackage{color}
\usepackage{amsmath}
\usepackage{amsxtra}
\usepackage{amstext}
\usepackage{amssymb}
\usepackage{latexsym}
\usepackage{dsfont}
\begin{document}

\title{Measuring the quantum state of a nanomechanical oscillator}

\author{Swati Singh and Pierre Meystre}
\affiliation{B2 Institute, Department of Physics and College of Optical
Sciences\\The University of Arizona, Tucson, Arizona 85721}

\date{\today}

\begin{abstract}
We propose a scheme to measure the quantum state of a nanomechanical oscillator cooled near its ground state of vibrational motion. This is an extension of the nonlinear atomic homodyning technique scheme first developed to measure the intracavity field in a micromaser. It involves the use of a detector-atom that is simultaneously coupled to the cantilever via a magnetic interaction and to (classical) optical fields via a Raman transition. We show that the probability for the atom to be found in the excited state is a direct measure of the Wigner characteristic function of the nanomechanical oscillator. We also investigate the backaction effect of this destructive measurement on the state of the cantilever.
\end{abstract}

\pacs{85.85.+j, 42.50.Wk, 42.50.-p, 42.50.Pq, 42.50.Dv}

\maketitle

There has been much recent progress toward taking mechanical systems into the quantum regime. Systems as diverse as nanomechanical resonators~\cite{Schwab04}, mirrors~\cite{gigan2006,Corbitt07}, micro-cavities~\cite{Kippenberg09} and nano-membranes~\cite{Harris08} are being cooled increasingly close to their quantum mechanical ground state. As a consequence of these developments, we can look forward to a number of novel applications ranging from the measurement of weak forces to tests of quantum mechanics, and from the development of a variety of quantum sensors to new applications in coherent control. Clearly, many of these applications rely on our ability not just to operate these systems in the quantum regime, but also to measure and control their quantum mechanical state.

Typical cantilever frequencies are in the KHz to MHz range, and the challenge is to characterize a phononic state in this frequency range. In the absence of phonon counters, other methods must be identified. A related difficulty arises in microwave cavity QED, where the absence of photon counters results in the need to develop alternative methods such as nonlinear atomic homodyning \cite{WilkensPM91} and Wigner tomography \cite{Wodkiewicz1,Wodkiewicz2,Davidovich,Haroche2,Haroche,Martinis} to characterize the state of the intracavity field. A common feature of these methods is that they consist of a series of destructive measurements that result in the reconstruction of the Wigner function of the field. In particular, Ref.~\cite{WilkensPM91} showed that the time-dependence of the upper state population of a two-state atom coupled to both the field to be characterized and a (classical) external field yields a direct measurement of the Wigner function of that field.

\begin{figure}[t]
\includegraphics[width=0.5 \textwidth]{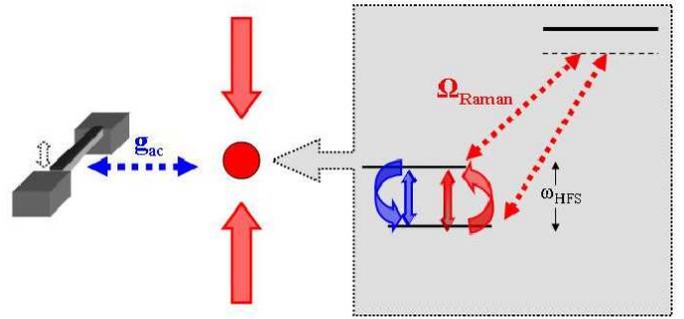}
\caption{(Color online). Arrangement
considered for the coupling a two level atom to a nanomechanical oscillator and optical field. A ferromagnetic domain enables coupling between the oscillator and the atomic levels with coupling strength $g_{ac}$. The same levels are also coupled via a Raman transition with a coupling strength $\Omega_{Raman}$.}
\label{fig:schematic}
\end{figure}

We show in the following that it is possible to extend this concept to the characterization of phononic fields, thereby providing a detection method to measure the quantum state of a micromechanical oscillator. Our scheme relies on the fact that it is considerably easier to measure the state of excitation of an atom than it is to directly measure the state of a field. It proceeds by transferring the state of the mechanical oscillator to the atomic upper state population, which can then easily be probed via a standard destructive measurement. One difficulty in achieving this goal is the poor frequency match between micromechanical and optical frequencies. One way to circumvent this mismatch is through the use of a Raman process to couple the upper and lower states of a two-level system with energy separation close to the cantilever frequency.

Recent progress in the nanofabrication of high frequency, high-Q resonators~\cite{RoukesNL07, RoukesJAP04} and their successful cooling~\cite{Schwab04,Rocheleau09} to extremely low thermal occupation numbers make them a viable candidate to demonstrate our scheme. The specific system that we have in mind is a doubly clamped nanomechanical resonator. The cantilever is magnetically coupled to a two-level atom through a magnetic domain on the cantilever. Good atomic candidates include alkali with hyperfine splitting close to cantilever's vibration frequency such as $^{6}$Li, with $\omega_{\rm HFS}$=228MHz, or $^{23}$Na, with $\omega_{\rm HFS}$=1.77GHz.

The mode of vibration of the cantilever to be measured is described as a simple harmonic oscillator of effective mass $m_c$ and frequency $\omega_c$,
\begin{equation}
\label{eq:NMC}
H_c=\hbar \omega_{c} \left(c^{\dagger}c+\frac{1}{2}\right),
\end{equation}
where $c$ and $c^{\dagger}$ are bosonic annihilation and creation operators.
The quantization axis $z$ of the two-state system, with Hamiltonian $H_a=\hbar \omega_0 \sigma_z/2 $, is chosen orthogonal to the direction of motion $x$ of the cantilever, so that the Zeeman coupling of the atom to the cantilever is given in the rotating wave approximation by \cite{haensch07}:
\begin{equation}
 \label{eq:Zeeman}
H_{ac}=\mu_{B}g_Fm_{Fx}G_{B} x_c
\end{equation}
where $\mu_B$ is the Bohr magneton, $g_F$ is the Land{\' e} factor of the $F$ quantum number of the atom, $m_{Fx}$ the projection of $F$ in the $x$-direction, $G_B$ the gradient of the magnetic field experienced by the atom, and $x_c$ is the position operator of the cantilever,
 \begin{equation}
\label{eq:xc}
    x_c=\sqrt{\frac{\hbar}{2 \omega_c m_c}}(c+c^\dagger).
\end{equation}
Such a coupling between a mechanical resonator and alkali atom has been demonstrated experimentally in Ref.~\cite{Kitching06}.

At resonance, $\omega_c=\omega_0$, and in the rotating wave approximation, the interaction between the cantilever and the atom is described by the Jaynes-Cummings Hamiltonian
\begin{equation}
 \label{acJCintxn}
H_{ac}=\hbar g_{ac}(c^\dagger \sigma_{-}+\sigma_{+}c)
\end{equation}
where
\begin{equation}
g_{ac}=\mu_{B}g_Fm_{Fx}G_{B} \sqrt{\frac{\hbar}{2 \omega_c m_c}}.
\end{equation}

Approximating the ferromagnetic domain as a magnetic dipole results in the  magnetic field gradient 
$$
|G_B|=3\mu_0|\mu_c|/4\pi r^4,
$$
where $\mu_c$ is the dipole moment of the resonator magnet. The strong dependence of $G_B$ on $r$, the distance between the atom and the cantilever, enables us to achieve very strong gradients close to the cantilever. It also provides us with a wide range of tunability in the interaction strength. When combined with a tunable splitting of various $m_F$ levels via a static magnetic field in the $z$-direction, it can result in a strong, resonant coupling for an experimentally reasonable range of parameters. In the case of optically trapped atoms, possible hyperfine transitions include collisionally stable stretched states such as the $|f=1/2,m_f=1/2\rangle$ and $|f=3/2,m_f=3/2\rangle$ sublevels of $^6$Li or the $|f=1,m_f=1\rangle$ and $|f=2,m_f=2\rangle$ states of $^{23}$Na. Alternatively, for atom chip traps one could consider magnetically trappable states such as $|f=1/2,m_f=-1/2\rangle$ and $|f=3/2,m_f=1/2\rangle$ in $^6$Li or $|f=1,m_f=-1\rangle$ and $|f=2,m_f=1\rangle$ in $^{23}$Na.

The coupling of the cantilever to a two-state system is not sufficient to determine its state, since in the Jaynes-Cummings interaction the state of the system depends only on correlation functions of the phononic field of the form
$\langle (c^\dagger c)^n \rangle$, $\langle (c^\dagger c)^n c \rangle$ or $\langle (c^\dagger c)^n c^\dagger \rangle$, where $n$ is an integer. To fully characterize it, we need instead access to a full set of correlation functions of the generic form $\langle c^\dagger c^\dagger \ldots c^\dagger cc \ldots c\rangle$. As discussed in Ref.~\cite{WilkensPM91} this can be achieved by coupling the atom to an additional external field. To account for the frequency mismatch between optical and hyperfine transition frequencies, in the present case that additional coupling is provided by a Raman transition involving a virtual transition to an additional excited state $|i\rangle$ as shown in Fig.~\ref{fig:schematic}. That state is coupled via electric dipole interaction to the lower state $|g\rangle$ by a far detuned classical field of Rabi frequency $\Omega_{L}=dE_L/\hbar$, where $d$ is the electric dipole moment of the transition, and $E_L$ is the electric field amplitude. The state $|i\rangle$ is also coupled to the other ground state $|e\rangle$ by a quantized field described by the bosonic annihilation and creation operators $a_k$ and $a_k^\dagger$. The vacuum Rabi frequency of that transition is $\Omega_k= d\sqrt{\omega_k/2\epsilon_0\hbar V}$.

Adiabatically eliminating the upper state $|i\rangle$, this Raman process is described by the effective Hamiltonian \cite{UysPM08}
\begin{equation}
H_{\rm Raman}=-\hbar \frac{\Omega_L \Omega_k}{ \delta_L} (a_k\sigma_+ + a_k^\dagger \sigma_-),
\end{equation}
where $\delta_L \gg \omega_0$ is the detuning of the two optical transitions involved in the process.

The situation is particularly simple if the coupling strength of the atom to the cantilever and the Raman fields are equal,
\begin{equation}
g_{ac} = -\frac{\Omega_L \Omega_k}{ \delta_L} \equiv g,
\end{equation}
a condition that can be realized either by adjusting the strength of the classical Raman field, or the Raman detuning, or the distance between the atom and the cantilever.

The interaction Hamiltonian reduces then to
\begin{equation}
\label{eq:totalH}
 H=\hbar \sqrt{2} g(\sigma_{-}A^{\dagger} +\sigma_{+}A)
\end{equation}
where we have introduced the bosonic operator
\begin{equation}
A=\frac{1}{\sqrt{2}}(a_k+c)
\end{equation}
with $[A, A^\dagger] = 1$. The Hamiltonian (\ref{eq:totalH}) is again a Jaynes-Cummings hamiltonian, but in terms of the phononic-photonic composite mode $a_k+c$. The key point here is that the correlation functions involved in the atomic evolution are now of the form $\langle (A^\dagger A)^n \rangle$, $\langle (A^\dagger A)^n A \rangle$ or $\langle (A^\dagger A)^n A^\dagger \rangle$, and it is the appearance of composite modes in these correlation functions that permit access to all correlation functions of the cantilever. At first sight, this might appear to raise the question of conservation of energy, since the photon energy $\hbar \omega_k$ is vastly higher than the phonon energy $\hbar \omega_c$. The point is that while not explicitly apparent when one of the optical fields is treated classically, at the microscopic level the emission (absorption) of a photon in mode $k$ is always accompanied by the absorption (emission) of energy by the other Raman field, and it is the energy difference $\hbar \omega_0$ between these two processes that is relevant.

The Jaynes-Cummings dynamics is well known. For an atom in the initial mixture
\begin{equation}
\rho_{\rm atom}=\rho_e|e\rangle\langle e|+\rho_g|g \rangle\langle g|
\end{equation}
and initially uncorrelated atom, cantilever, and optical fields, $\rho(0)=\rho_a(0)\otimes \rho_c(0)\otimes \rho_o(0)$,
the probability $P_e$ to find the atom in the excited state at time $\tau$ is
\begin{eqnarray}
\label{eq:ProbEx}
P_e &=&\frac{1}{2}+\frac{1}{2}\rho_e\langle\cos{(2\sqrt{2}g\tau\sqrt{A^{\dagger}A+1})}\rangle\nonumber \\
& &-\frac{1}{2}\rho_g\langle\cos{(2\sqrt{2}g\tau\sqrt{A^{\dagger}A+1})}\rangle
\end{eqnarray}
where $\langle X\rangle \equiv Tr\left (\rho_c(0)\rho_o(0) X \right ).$

Considerable insight can be gained by considering the situation where the cantilever state is close to the ground state, with $\langle n_c\rangle \equiv \langle c^\dagger c \rangle \simeq 1$, a situation of much experimental interest, and the Raman field is in a coherent state $|\beta_k\rangle$ with $\langle \beta_k\rangle = \sqrt{I}\exp(i\phi)$ and $|\beta_k|^2 \equiv I \gg \langle n_c\rangle$. We then have
\begin{eqnarray}
\sqrt{A^{\dagger}A+1}&\approx&\sqrt{A^{\dagger}A}\nonumber \\
&=& \sqrt{\frac12 \left [\langle n_c\rangle + I + I^{1/2}(c^\dagger e^{i\phi}+ c e^{-i\phi}) \right ]} \nonumber \\
&\approx& \sqrt{\frac{I}{2}}+ \frac{1}{{2\sqrt 2}}\left( c^\dagger e^{i\phi} + c e^{-i\phi}\right)
\end{eqnarray}
and the excited state probability simplifies to
\begin{equation}
P_e=\frac{1}{2}+\frac{1}{4}\left(\rho_{\uparrow}-
\rho_{\downarrow}\right)\left(e^{2ig\tau\sqrt{I}}C_W(\mu)+h.c.\right)
\end{equation}
where
\begin{equation}
\mu=ig\tau \exp(i\phi)
\end{equation}
and $C_W$ is the Wigner characteristic function of the cantilever phonon mode,
\begin{equation}
C_W(\mu)=Tr(\rho_{c}e^{\mu c^{\dagger}-\mu^*c}).
\end{equation}
As is the case in Wigner tomography, the full Wigner characteristic function, and hence the Wigner function of the phonon mode, can be reconstructed by varying the interaction time $\tau$ and/or the phase $\phi$ of the Raman field. We emphasize that this is a destructive measurement scheme, and that in general a large sequence of measurements that scan $\mu$ in the complex plane starting from identical initial conditions are necessary to reconstruct the state of the cantilever. This is similar to the situation in cavity QED and in circuit QED~\cite{Haroche2,Martinis}.

Indeed, the back-action of a sequence of repeated measurements of the kind proposed here on the state of the cantilever is normally very significant. To illustrate this point we assume for concreteness that a series of measurements on the cantilever are performed by a sequence of detector atoms initially in their ground state. Their ground-state population is measured after an interaction time $\tau$. It is easy to show that for classical Raman fields and in the absence of dissipation, a cantilever initially in a pure state remains then in a pure state. We concentrate on this simple case for clarity, noting that the extension to mixed states is straightforward.

\begin{figure}[ht]
\begin{center}
$\begin{array}{cc}
\mbox{\bf (a)} &
\includegraphics[width=2.75in]{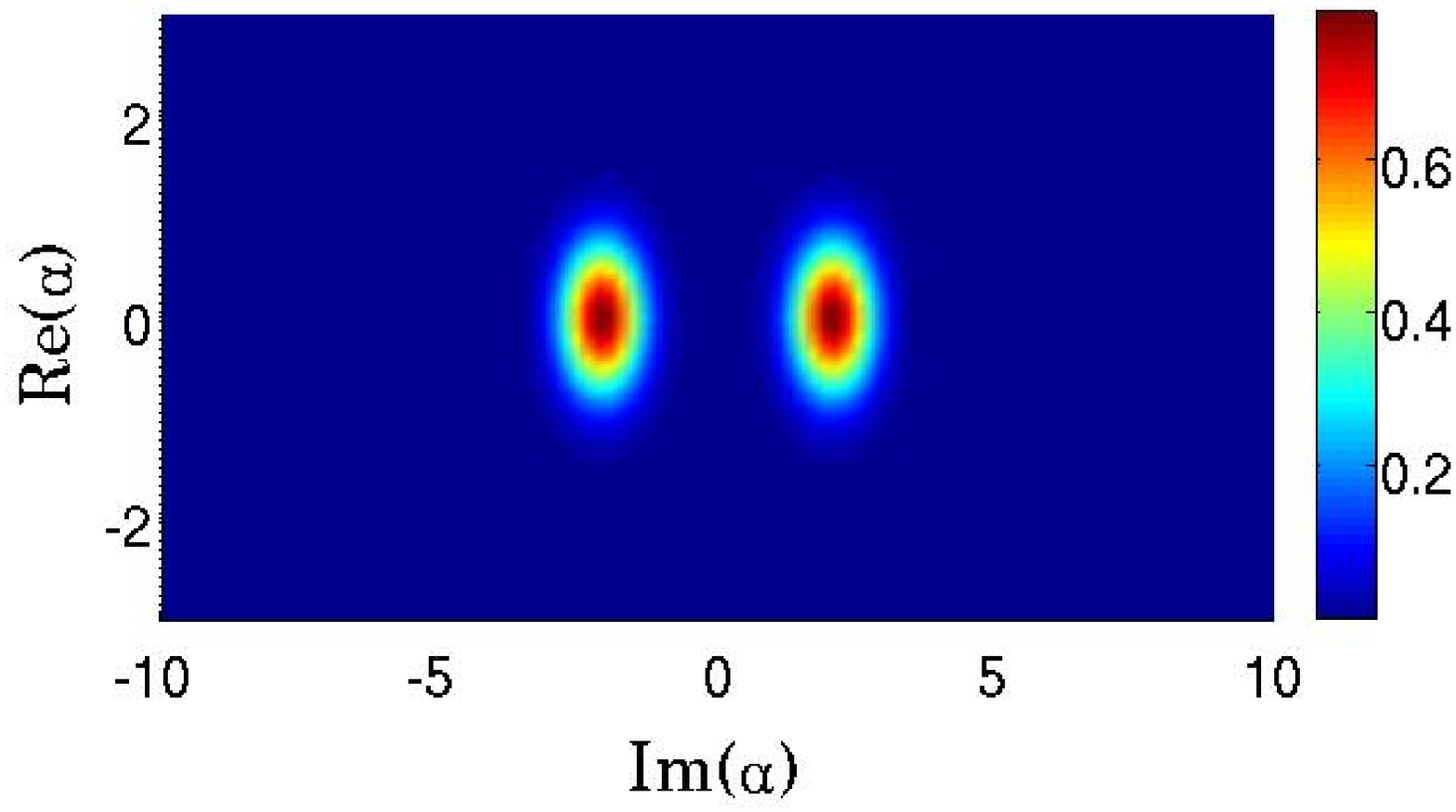} \\[0.0cm]
\mbox{\bf (b)} &
\includegraphics[width=2.75in]{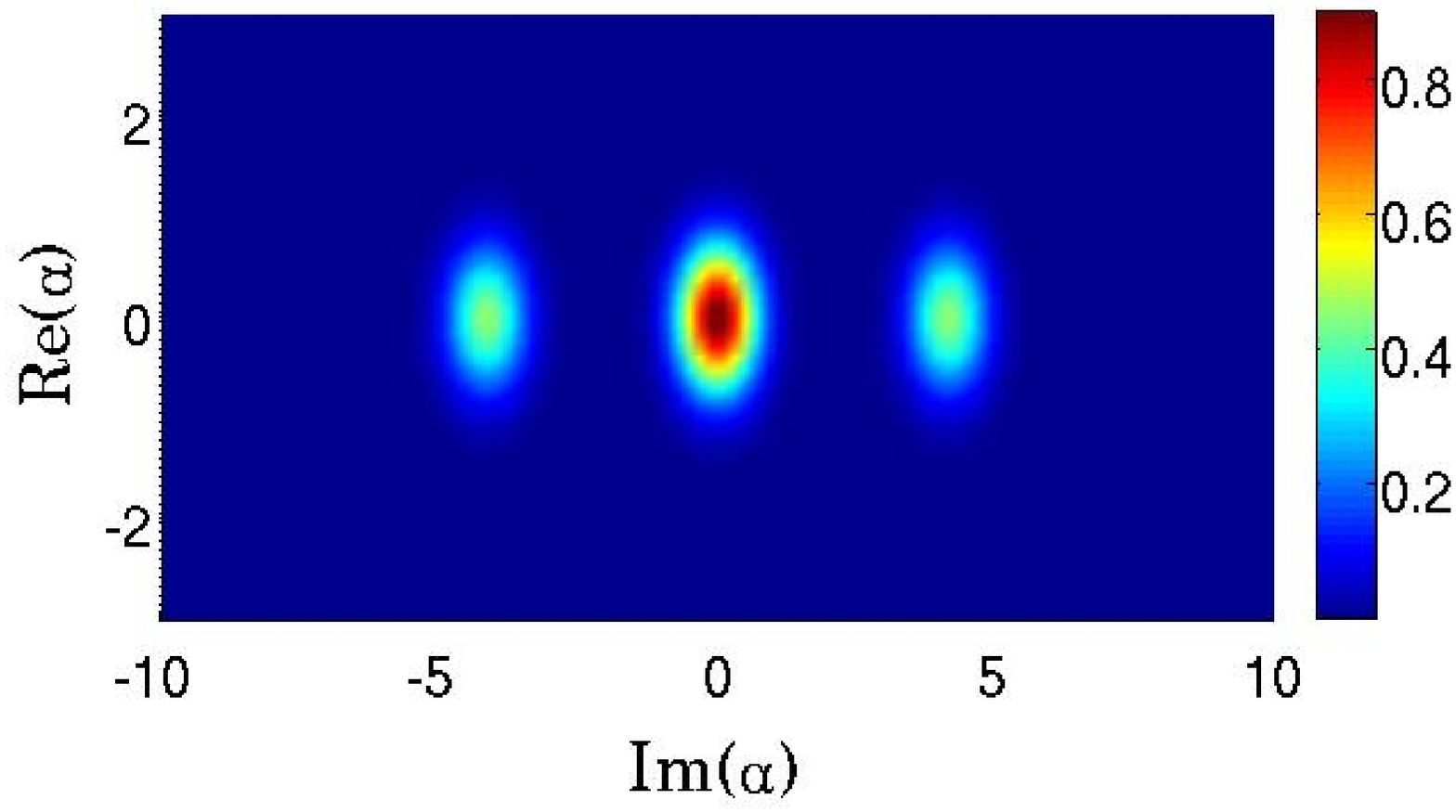} \\ [0.0cm]
\mbox{\bf (c)} &
\includegraphics[width=2.75in]{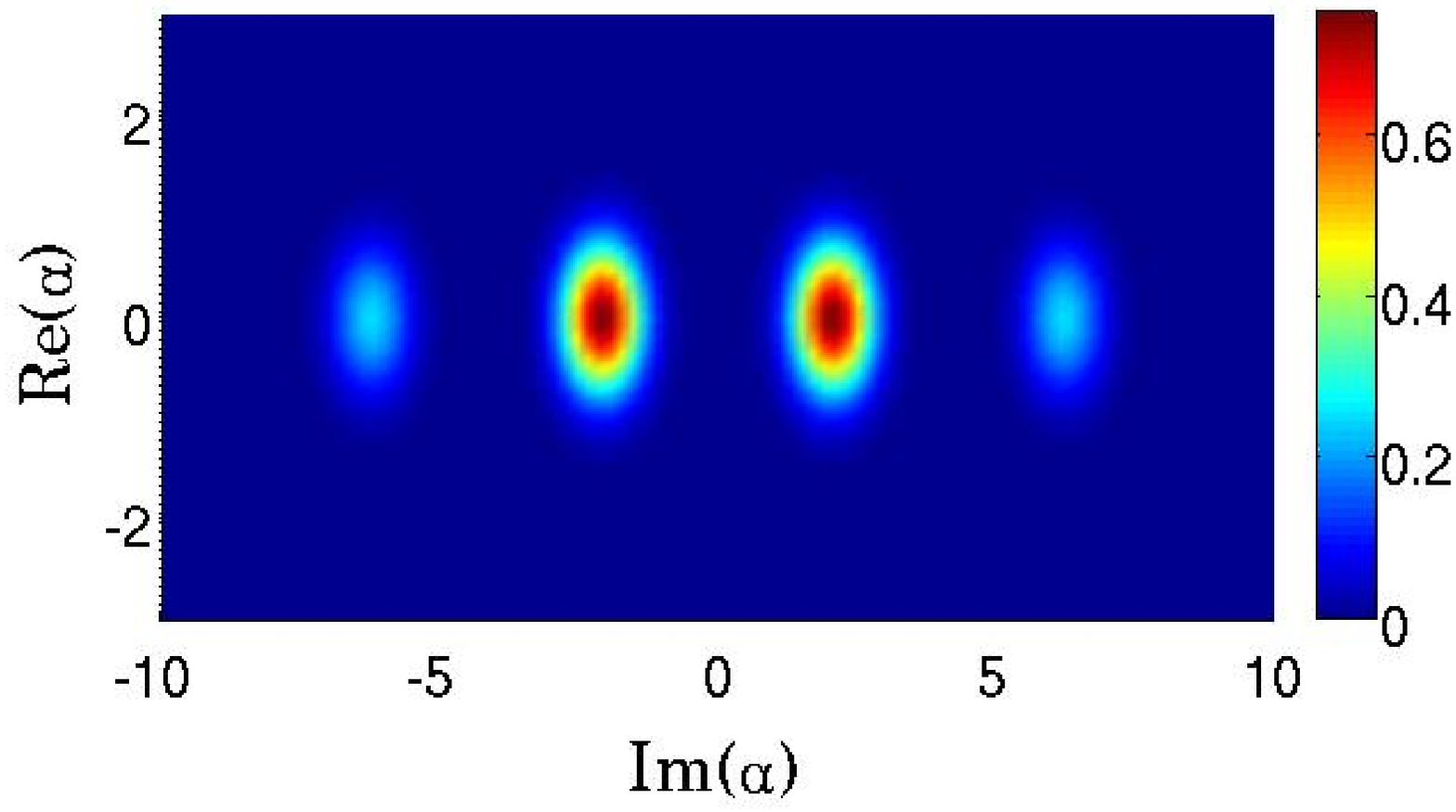} \\ [0.0cm]
\mbox{\bf (d)} &
\includegraphics[width=2.75in]{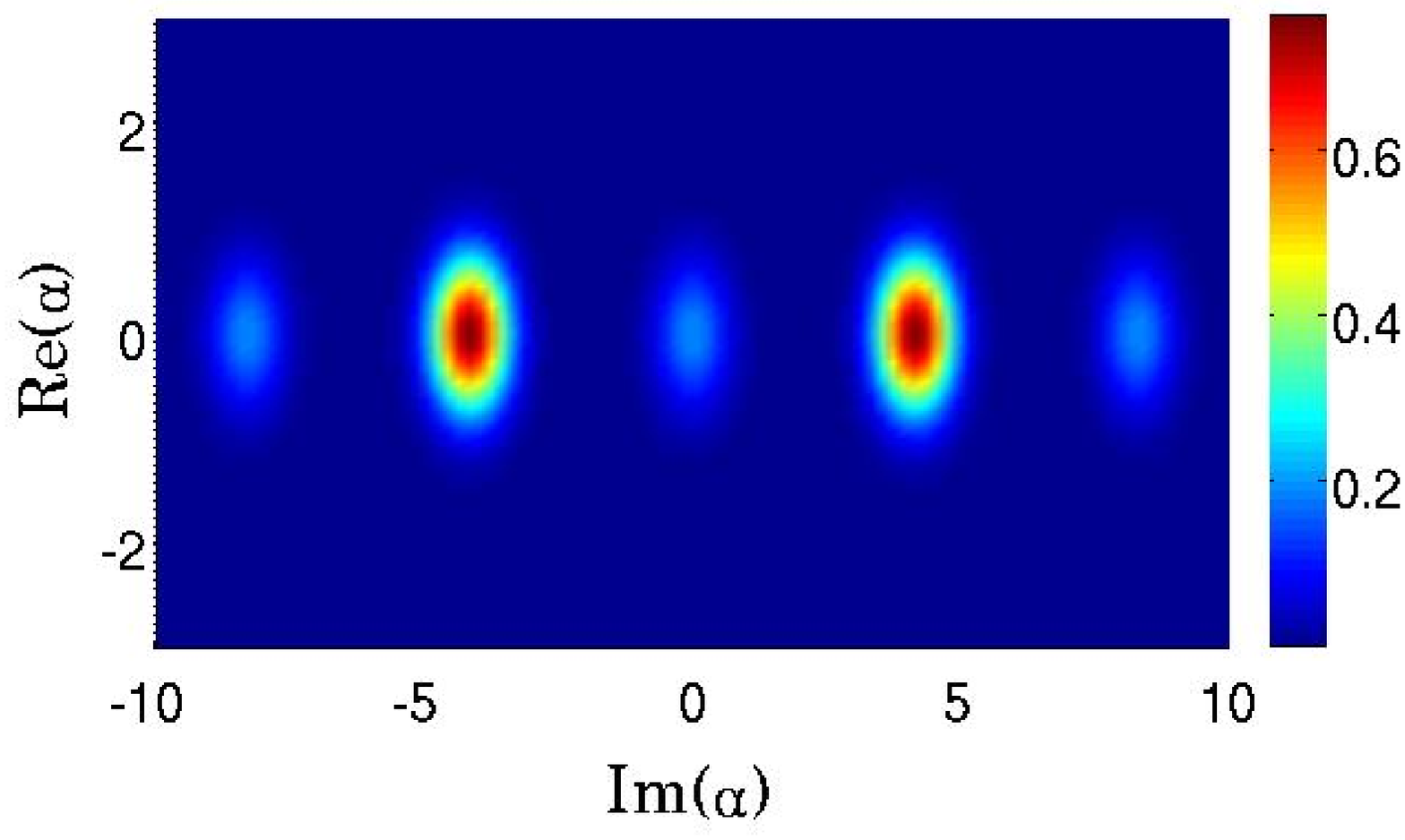} \\ [0.0cm]
\end{array}$
\end{center}
\caption{(Color online) Wigner distribution function of the cantilever after (a) a first measurement after time $\tau$=5ms of evolution, and (b, c, d) subsequent measurements (b,c,d) of equal duration $\tau$. Here the cantilever was taken to be initially in the ground state, and $\mu$ is purely imaginary $\mu$=4.15$i$.}
\label{fig:measureBaction}
\end{figure}

Immediately following the $i$-th measurement, the state of the cantilever is given by~\cite{ZauggPM93}
\begin{equation}
 |\psi_{c,i}\rangle = \frac{\langle g|U(\tau)|\psi_{c,i-1};g\rangle} {\sqrt{P_{g}(\tau)}}
\end{equation}
where $U(\tau)$ is the evolution operator associated with the Jaynes-Cummings Hamiltonian~(\ref{eq:totalH}) and $|\psi_{i}; g\rangle$ is the state of the full system just as the $i$-th detector atom is switched on. In the limiting case $I \gg \langle n_c\rangle$ this yields
\begin{eqnarray}
|\psi_{c,i}\rangle=\frac{1}{2\sqrt{P_{g}}}\left(\cos(g\tau\sqrt{I})\left[D(\mu/2)+D(\mu/2)\right]\right)+\nonumber\\
\frac{1}{2\sqrt{P_{e}}}\left(i\sin(g\tau\sqrt{I})\left[D(\mu/2)-D(-\mu/2)\right]\right)|\psi_{c,i-1}\rangle,
\end{eqnarray}
where $D(\mu)=\exp(\mu c^\dagger - \mu^* c)$ is the displacement operator. At the beginning of the $(i+1)$-th measurement, the initial state of the full system is therefore $|\psi_{c,i}; g\rangle$ (neglecting dissipation and the free evolution between atoms), from which the next iteration is started.
The successive measurements therefore displace the state of the cantilever into different regions of phase space, and the distance between these regions keeps increasing as a function of time. This is illustrated in Fig. \ref{fig:measureBaction} that shows the Wigner Function of the cantilever after successive measurements done at equal time intervals, for a cantilever intially in vacuum state.

In summary, we have proposed a destructive measurement scheme to determine the Wigner function of a mechanical cantilever cooled near its ground state of vibration. Our proposed setup involves a detector atom coupled to the phonon mode of relevance, and to a pair of optical fields that induce a Raman transition between the ground and excited state of the detector atom. This scheme is an extension of a method previously considered for the detection of quantized microwave fields to the case of phonons detection. We have also proposed a realistic set of experimental parameters for which a demonstration experiment should be feasible.

We conclude by remarking that the same coupling scheme can also be used to prepare an arbitrary quantum state of the cantilever. This can be seen by a straightforward extension of the results of Ref.~\cite{LawEberly96}, which demonstrated that a two-level system coupled to a classical and a quantum field can be used to generate an arbitrary state of that field, provided the two couplings can be tuned independently. An experimental realization of such quantum states was recently demonstrated in a circuit-QED system by Hofheinz et al.~\cite{Martinis}. By independently controlling the Raman and magnetic coupling, our system can likewise also provide the ability to generate arbitrary quantum states of the cantilever.

We thank K. Zhang, W. Chen, D. Goldbaum and M. Bhattacharya for stimulating discussions. This work is supported in part by the US Office of Naval Research, by the National Science Foundation, and by the US Army Research Office.


\begin{thebibliography}{10}

\bibitem{Schwab04}
M.D. LaHaye, O. Buu, B. Camarota, K.C. Schwab, Science {\bf 304}, 74 (2004).

\bibitem{gigan2006}
S. Gigan \textit{et al.}, Nature {\bf 444},  67  (2006).

\bibitem{Corbitt07}
T. Corbitt \textit{et al.}, Phys. Rev. Lett. {\bf 98},  150802  (2007).

\bibitem{Kippenberg09}
A. Schliesser \textit{et al.}, Nature Physics {\bf 5}, 509 (2009).

\bibitem{Harris08}
J.D. Thompson \textit{et al.}, Nature {\bf 451}, 72 (2008).

\bibitem{WilkensPM91}
M. Wilkens and P. Meystre, Phys. Rev. A. {\bf 43}, 3832  (1991).

\bibitem{Wodkiewicz1}
K. Banaszek and K. Wodkiewicz, Phys. Rev. Lett. {\bf 76}, 4344 (1996).

\bibitem{Wodkiewicz2}
K. Banaszek, C. Radzewicz, K. Wodkiewicz and J. S. Ksasinski, Phys. Rev. A {\bf 60}, 674 (1999).

\bibitem{Davidovich}
L. G. Lutterbach and L. Davidovich, Phys. Rev. Lett. {\bf 78}, 2547 (1997).

\bibitem{Haroche2}
S. Del{\' e}glise {\it et al}, Nature {\bf 445}, 510 (2008)

\bibitem{Haroche}
S. Haroche and J. M. Raimond, ``Exploring the quantum -- atoms, cavities and photons'', Oxford Univ. Press (2006).

\bibitem{Martinis}
M. Hofheinz1 {\it et. al.}, Nature {\bf 459}, 546 (2009).

\bibitem{Rocheleau09}
T. Rocheleau, T. Ndukum, J. Hertzerg and K. Schwab, Private Communication (2009).

\bibitem{RoukesNL07}
X. L. Feng, R. He, P. Yang and M. L. Roukes, Nano Lett. {\bf 7},1953 (2007).

\bibitem{RoukesJAP04}
K.L. Ekinci, Y.T. Yang and M.L. Roukes, J. Appl. Phys. {\bf 95}, 2682 (2004).

\bibitem{haensch07}
P. Treutlein  \textit{et. al}, Phys. Rev. Lett. {\bf 99},  140403  (2007).

\bibitem{Kitching06}
Ying-Ju Wang \textit{et. al}, Phys. Rev. Lett. {\bf 97}, 227602 (2006).

\bibitem{UysPM08}
H. Uys and P. Meystre, Phys. Rev. A. {\bf 77}, 063614 (2008).

\bibitem{Savard99}
T. Savard \textit{et. al}, Phys. Rev. A. {\bf 60}, 4788 (1999).

\bibitem{barnett}
S.M. Barnett and P.M. Radmore, ``Methods in Theoretical Quantum Optics'', Oxford Univ. Press (1997).

\bibitem{RoukesAPL96}
A. N. Cleland and M. L. Roukes, Appl. Phys. Lett. {\bf 69}, 2653 (1996).

\bibitem{ZauggPM93}
T. Zaugg, M. Wilkins and P. Meystre, Found. Phys. {\bf 23}, 857 (1993).

\bibitem{LawEberly96}
C. K. Law and J. H. Eberly, Phys. Rev. Lett. {\bf 76}, 1055 (1996).

\end{thebibliography}
\end{document}